\documentclass[aps,prl,twocolumn,showpacs,amsmath,amssymb,amsfonts,nofootinbib,long]{revtex4}
\usepackage{epsfig,latexsym}

\newcommand\G{\mbox{G}}
\newcommand\Mpc{\mbox{Mpc}}
\newcommand\kpc{\mbox{kpc}}
\newcommand\pc{\mbox{pc}}
\newcommand\GeV{\mbox{GeV}}

\newcommand\Gyr{\mbox{Gyr}}
\newcommand\km{\mbox{km}}
\renewcommand\sec{\mbox{sec}}

\newcommand\lambdakpc{\lambda_{\mbox{\scriptsize 10kpc}}}
\newcommand\tot{\mbox{\scriptsize tot}}
\newcommand\m{\mbox{\scriptsize min}}

\newcommand\Pl{\mbox{\scriptsize Pl}}

\newcommand\GH{\mbox{\scriptsize GH}}

\begin{document}

\title{Inflation-Produced Magnetic Fields in Nonlinear Electrodynamics}

\author{L. Campanelli$^{1,2,3,4}$}
\email{leonardo.campanelli@ba.infn.it}
\author{P. Cea$^{3,4}$}
\email{paolo.cea@ba.infn.it}
\author{G.L. Fogli$^{3,4}$}
\email{gianluigi.fogli@ba.infn.it}
\author{L. Tedesco$^{3,4}$}
\email{luigi.tedesco@ba.infn.it}

\affiliation{$^1$Dipartimento di Fisica, Universit\`{a} di Ferrara, I-44100 Ferrara, Italy}
\affiliation{$^2$INFN - Sezione di Ferrara, I-44100 Ferrara, Italy}
\affiliation{$^3$Dipartimento di Fisica, Universit\`{a} di Bari, I-70126 Bari, Italy}
\affiliation{$^4$INFN - Sezione di Bari, I-70126 Bari, Italy}

\date{October, 2007}


\begin{abstract}
We study the generation of primeval magnetic fields during
inflation era in nonlinear theories of electrodynamics. Although
the intensity of the produced fields strongly depends on
characteristics of inflation and on the form of electromagnetic
Lagrangian, our results do not exclude the possibility that these
fields could be astrophysically interesting.
\end{abstract}


\maketitle


\section{I. Introduction}

All galaxies seem to be permeated by magnetic fields with
intensities of order $B_{\rm galactic} \simeq 10^{-6} \G$
\cite{Beck}.

To explain the galactic magnetism, generally one needs the
presence of seed magnetic fields prior to protogalaxy collapse.
When a protogalaxy collapses to form a galactic disk, magnetic
fields suffer an amplification (mainly due to magnetic flux
conservation) of order $\mathcal{A}_{\rm pg} \simeq 10^4$
\cite{Lesch}. Moreover, due to magnetohydrodynamic turbulence
effects and differential rotation of galaxy, seed fields can be
further amplified. This last mechanism, know as ``galactic
dynamo'' \cite{Dynamo}, can be very efficient and, in principle,
allow extremely week seeds to reproduce the properties of
presently-observed galactic fields. Dynamo action produces an
exponential amplification $\mathcal{A}_{\rm dyn} \simeq e^{\Gamma
\Delta t}$, where the growth rate $\Gamma$ is a model-dependent
quantity, and $\Delta t = t_f - t_i$ is the time during which the
dynamo operates. The minimum and maximum values of $\Gamma$ that
can be found in the literature are $\Gamma \simeq 0.45 \Gyr^{-1}$
\cite{Gamma045} and $\Gamma \simeq 5 \Gyr^{-1}$ \cite{Gamma5}. In
a spatially-flat Friedman universe, the time interval is
\cite{Widrow}
$\Delta t = (2/3H_0\Omega_{\Lambda}^{1/2}) \ln [(\omega_f +
\sqrt{1+\omega_f^2})/(\omega_i + \sqrt{1+\omega_i^2})]$.
Here, $H_0 = 100 h \, \km \, \sec^{-1} \Mpc^{-1}$ is the Hubble
constant ($h \simeq 0.73$ \cite{WMAP3yr}), $\omega_{i,f} =
(1+z_{i,f})^{-3/2} (\Omega_{\Lambda}/\Omega_m)^{1/2}$, $z$ is the
red-shift, and $\Omega_m \simeq 0.28$ and $\Omega_{\Lambda} \simeq
0.72$ \cite{WMAP3yr} are the actual energy densities, in units of
the critical density $\rho_{\rm cr} = 3 H_0^2/8 \pi G$, associated
to matter and cosmological constant.
Un upper bound on the value of red-shift at which the dynamo
begins to operate, $z_i$, is given by the red-shift $z_{\rm pg}
\simeq 50$ at which a protogalaxy separates from the Hubble flow
to then collapse \cite{Widrow}. If dynamo is efficient during
galactic disk formation or if its amplification becomes effective
only after that, is still not clear. In the latter case, since
astronomical observations indicate that disk galaxies at $z \simeq
3$ are still in progress of being formed \cite{z3}, one should
take this value of red-shift as a conservative bound on $z_i$.
A lower bound on $z_f$ is $z_f \simeq 0.4$, since microgauss
magnetic fields have been detected in galaxies at that red-shift
\cite{z04}.

The galactic magnetism can be then explained as the result of the
amplification of comoving seed fields as strong as $B_{\rm seed}
\gtrsim 10^{-6} (1+z_{\rm pg})^{-2} \mathcal{A}_{\rm pg}^{-1}
\mathcal{A}_{\rm dyn}^{-1} \G$, where the factor $(1+z_{\rm
pg})^{-2}$ takes into account the adiabatic scaling of the
magnetic field from the protogalaxy collapse until today. Taking
$(\Gamma,z_i,z_f) = (5,50,0.4)$ we have $B_{\rm seed} \gtrsim
10^{-33} \G$, while for $(\Gamma,z_i,z_f) = (0.45,3,0.4)$ we get
$B_{\rm seed} \gtrsim 10^{-15} \G$.
In order to have an efficient galactic dynamo, however, the seed
magnetic field must be correlated on comoving scales of order $10
\kpc$.
\\
We observe, also, that without dynamo amplification a comoving
seed field as strong as $B_{\rm seed} \gtrsim 10^{-14} \G$ is
needed to explain galactic magnetism. In this case, however, the
field must be correlated on comoving scales of order of linear
dimensions of a protogalaxy, that is $1 \Mpc$.

Surprisingly, nanogauss magnetic fields correlated on scales of
order $1 \Mpc$, have been also detected in galaxy clusters and
superclusters. This last observation seems to indicate that the
entire universe is magnetized \cite{Widrow,Magnetic,Vallee}.

Essentially, there are two possible classes of mechanisms to
produce cosmic fields depending on when they are generated
\cite{Dolgov}: Astrophysical mechanisms acting during large-scale
structure formation \cite{Generation1}, and mechanisms acting in
the primordial universe, during
\cite{Turner,Ratra,Lemoine,Garretson,Anber,Gasperini,Dolgov1,Generation2}
or before \cite{Generation3} inflation.
However, we can admit the existence of strong fields in the
primordial universe provided that their presence do not spoil
predictions of the standard cosmological model, such as that of
Big Bang Nucleosynthesis (BBN) \cite{BBN}, Large-Scale-Structure
Formation (LSS) \cite{LSS}, and Cosmic Microwave Background (CMB)
\cite{CMB}. It turns out that limits coming from LSS and CMB are
more stringent than those from BBN. Putting together the limits
found in Refs.~\cite{LSS,CMB} it results that, for comoving scales
in the range $400\pc \lesssim \lambda \lesssim d_H(t_0)$, where
$d_H(t_0) \sim H_0^{-1} \sim 4000\Mpc$ is the present size of the
universe, the maximum strength allowed to comoving primordial
fields is $B \sim 10^{-9} \G$.

In the ambit of generation of cosmological fields in the early
universe, the mechanisms operating during inflation are
particularly attractive since they produce large-scale correlated
fields. Magnetic fields created after inflation, instead, suffer
from a ``small-scale problem'', that is their comoving correlation
length is much smaller then the characteristic scale of the
observed cosmic fields [however, if magnetohydrodynamic turbulence
operates during their evolution, an enhancement of correlation
length can occur (see, e.g., Ref.~\cite{MHD})].

It is worth noting that, due to conformal invariance of standard
(Maxwell) electrodynamics and to the fact the spacetime described
by the Robertson-Walker metric is conformally flat, magnetic
fields generated at inflation are vanishingly small. For this
reason, all generating models proposed in the literature repose on
the breaking of conformal invariance of Maxwell theory. This has
been attained, for instance, by non-minimally coupling the photon
with gravity \cite{Turner}, introducing interactions of photons
with scalar, pseudoscalar, or vector fields (such as inflaton
\cite{Ratra}, dilaton \cite{Lemoine}, pseudo-Goldstone bosons
\cite{Garretson}, axion \cite{Anber}, or ``graviphoton''
\cite{Gasperini}), taking into account the so-called Quantum
Conformal Anomaly \cite{Dolgov1}, and so on \cite{Generation2}.

In this paper, we study the possibility to generate seed magnetic
fields during inflation in nonlinear theories of electrodynamics
(NLE) described by the general action
\begin{equation}
\label{action} S = \frac{1}{4 \pi} \int \! d^4x \sqrt{-g} \,
\mathcal{L}(F),
\end{equation}
where $F = \frac14 F_{\mu \nu} F^{\mu \nu}$, with $F_{\mu \nu} =
\partial_{\mu} A_{\nu} - \partial_{\nu} A_{\mu}$ the
electromagnetic field strength tensor, and $g = \det||g_{\mu
\nu}||$ is the determinant of the metric tensor.
\\
Since the standard Maxwell theory is a good theory for low
energies, we shall assume that nonlinear Lagrangians reduce to the
Maxwell one, $\mathcal{L}(F) \simeq -F$, in the limit of small
fields $F$.
\\
A considerable amount of interest has emerged in the last few
years in cosmological effects of nonlinear electrodynamics
\cite{NLE,Novello}.
This is due principally to the fact that some theories of NLE are
able to produce inflation \cite{NLEinfl,Novello}, a period of
cosmic acceleration \cite{NLEacc,Novello}, and can also avoid the
problem of initial singularity \cite{NLEsing,Novello,Lorenci}.
\\
In general, nonlinear electrodynamic theories are non-conformally
invariant. As we shall see in the next Section, depending on the
actual form of the Lagrangian, astrophysically interesting
magnetic fields can be generated during inflation.


\section{II. Generation of seed fields in NLE}

\subsection{A. Equations of Motion}

We will work in a flat universe described by a Robertson-Walker
metric, $ds^2 = a^2(\eta)(d\eta^2 - d{\textbf x}^2)$, where
$a(\eta)$ is the expansion parameter and $\eta$ is the conformal
time related to the cosmic time $t$ through $d \eta = dt/a$.
Introducing the electric and magnetic fields ${\textbf E}$ and
${\textbf B}$ in the usual way as $F_{0i} = -a^2 E_i, \;\;\;
F_{ij} = \epsilon_{ijk} a^2 B_k$ (Latin indices range from $1$ to
$3$), and varying the action (\ref{action}) with respect to
$A_{\mu}$, we get the equations of motion:
\begin{equation}
\label{Max2} \frac{\partial  (a^2 {\textbf E})}{\partial \eta} -
\nabla \times (a^2 {\textbf B}) = - \frac{\partial  \ln
\mathcal{L}_F}{\partial \eta} \, a^2 {\textbf E} + \nabla\ln
\mathcal{L}_F \times a^2 {\textbf B},
\end{equation}
and $\nabla \cdot (\mathcal{L}_F {\textbf E}) = 0$, together with
the Bianchi identities:
$\partial_\eta(a^2 {\textbf B}) + \nabla \times (a^2 {\textbf E})
= 0$
and $\nabla \cdot {\textbf B} = 0$. Here, subscript on
$\mathcal{L}$ denotes differentiation, and spatial derivatives are
taken with respect to comoving coordinates.
\\
We are interested to the evolution of electromagnetic fields
outside the horizon, that is to modes whose physical wavelength is
much greater then the Hubble radius $d_H$, $\lambda_{\rm phys} \gg
d_H$, where $\lambda_{\rm phys} = a \lambda$, $d_H \sim H^{-1}$,
and $\lambda$ is the comoving wavelength. Since $\eta a \sim
H^{-1}$, introducing the comoving wavenumber $k = 2\pi/\lambda$,
the above condition reads $|k\eta| \ll 1$.
Observing that the first Bianchi identity gives
${\textbf B}(k\eta) \sim k\eta {\textbf E}(k\eta)$,
we have that ${\textbf B}^2$ is negligible with respect to
${\textbf E}^2$, and we can write $\partial_{\eta} \! \ln
\mathcal{L}_F \simeq -(\partial_\eta {\textbf E}^2/2) \, (d\ln
\mathcal{L}_F/dF)$. Moreover, we can neglect the second term with
respect to the first one both in the left- and right-hand-side of
Eq.~(\ref{Max2}). In fact, it results
$|\nabla \times a^2 {\textbf B}|/|\partial_{\eta} (a^2 {\textbf
E})| \sim |k \eta|^2 \ll 1$, and
$|\nabla\ln \mathcal{L}_F \times a^2 {\textbf
B}|/|(\partial_{\eta} \! \ln \mathcal{L}_F) \, a^2 {\textbf E}|
\sim |k \eta|^2 \ll 1. $
Finally, multiplying Eq.~(\ref{Max2}) by ${\textbf E}$, and
solving with respect to $F \simeq -\frac12 {\textbf E}^2$, we get
\begin{equation}
\label{solution} (\mathcal{L}_F)^2 F \propto a^{-4}.
\end{equation}
Knowing the form of nonlinear Lagrangian (see below), from the
above equation one gets the evolution law for the electric field
outside the horizon. Consequently, using the first Bianchi
identity, one finds how super-horizon magnetic fields scale in
time during inflation (see subsection D).

\subsection{B. Initial Electromagnetic Spectrum}

During inflation, all fields are quantum mechanically exited.
Because the wavelength $\lambda$ associated to a given fluctuation
grows faster then the horizon, there will be a time, say $t_1$,
when this mode crosses outside the horizon itself. After that,
this fluctuation cannot collapse back into the vacuum being not
causally self-correlated, and then ``survives'' as a classical
real object \cite{Dimopoulos}. The energy associated to a given
fluctuation is subjected to the uncertainty relation, $\Delta E
\Delta t \gtrsim 1$. Therefore, the energy density in the volume
$\Delta V$, $\mathcal{E} = \Delta E/\Delta V$, is approximatively
given by $\mathcal{E} \sim H^4$, where $H$ is the Hubble
parameter. Here, we used the fact that at the horizon crossing
$\Delta t \sim H^{-1}$ and $\Delta V \sim H^{-3}$
\cite{Dimopoulos}.
When a comoving length $\lambda$ crosses the horizon it results
$|k\eta| \simeq 1$, and then from the first Bianchi identity we
get ${\textbf B}^2(\lambda) \simeq {\textbf E}^2(\lambda)$.
Therefore, since $F = -\frac12 ({\textbf E}^2-{\textbf B}^2)$, at
the crossing the nonlinear terms in the electromagnetic Lagrangian
are negligible, and the energy density is simply given by
$\mathcal{E} \simeq  (1/8\pi) \, ({\textbf E}^2 + {\textbf B}^2)$.
Consequently, the spectra of the quantum-mechanically generated
electric and magnetic fluctuations are given by
\begin{equation}
\label{BE1} \frac{{\textbf B}^2(\lambda)}{4\pi}|_{t_1} \sim
\frac{{\textbf E}^2(\lambda)}{4\pi}|_{t_1} \sim H^4|_{t_1} =
\left(\frac{8\pi}{3}\right)^{\!2}
\frac{\rho^2_{\tot}(\lambda)}{m_{\Pl}^4}|_{t_1},
\end{equation}
where in the last equality we used the Friedman equation $H^2 =
(8\pi/3) \rho_{\tot}/m_{\Pl}^2$. Here, $\rho_{\tot}$ is the total
energy density during inflation and $m_{\Pl} \simeq 10^{19} \GeV$
the Planck mass.
\\
In the following we shall consider both the case of de Sitter
inflation and the case of ``Power-Law Inflation''. It is useful to
write the expansion parameter as $a(\eta) \propto \eta^s$. During
de Sitter inflation $s=-1$ and the total energy is a constant,
\footnote{In de Sitter inflation, the spectrum of electromagnetic
fluctuations when crossing the horizon is $|A_\mu| \sim
|F_{\mu\nu}|/H \sim H$, that is a scale-invariant spectrum
corresponding to the Gibbons-Hawking temperature $T_{\GH} =
H/(2\pi)$ \cite{Gibbons}.}
$\rho_{\tot} \equiv M^4$.
For power-law inflation described by the equation of state
$p_{\tot} = \gamma \rho_{\tot}$ with $-1 < \gamma < 1/3$, we have
$s=2/(1+3\gamma)$ and the total energy decreases as a power of the
expansion parameter, $\rho_{\tot} \propto a^{-3(1+\gamma)}$.
Therefore, the total energy, when a comoving length $\lambda$
crosses outside the horizon, depends on $\lambda$ [see
Eq.~(\ref{outside})].

\subsection{C. Plasma Effects}

The effects of a conducting plasma in the early universe on the
evolution of magnetic fields, are taken into account by adding to
the electromagnetic Lagrangian the source term $J^{\mu} \!
A_{\mu}$ \cite{Turner}. Here, the external current $J^{\mu}$,
expressed in terms of the electric field, has the form
$J^{\mu} = (0, \sigma_c {\textbf E})$,
where $\sigma_c$ is the conductivity. Plasma effects introduce, in
the left-hand-side of Eq.~(\ref{Max2}), the extra term $-
a\sigma_c (a^2 {\textbf E})$. In the limit of high conductivity,
$\sigma_c \rightarrow \infty$, one finds ${\textbf E} \rightarrow
0$ and, consequently, from the first Bianchi identity, it follows
that the magnetic field is frozen into the plasma and evolves
adiabatically, $a^2 {\textbf B} \sim \mbox{const}$.
More precisely, the solution of Eq.~(\ref{Max2}), for modes
outside the horizon and when plasma effects are taken into
account, is
$(\mathcal{L}_F)^2 F \propto a^{-4} \exp(-2\!\int \!d\eta \,
a\sigma_c)$.
Observing that $\int \!d\eta \, a\sigma_c \sim \sigma_c/H$, we get
that the magnetic field can be considered as frozen into the
plasma when $\sigma_c \gg H$.
After inflation, the universe enters in the so-called reheating
phase, during which the energy of the inflaton is converted into
ordinary matter. In this paper, we shall restrict our analysis to
the case of instantaneous reheating, that is after inflation the
universe enters the radiation dominated era.
In this era, the conductivity is approximatively equal to
$\sigma_c \sim T$ \cite{Turner}, while $H \sim T^2/m_{\Pl}$
\cite{Kolb}, where $T$ is the temperature. Hence, the condition of
freezing of the magnetic field becomes $T \ll m_{\Pl}$.
\\
The spectrum of gravitational waves generated at inflation is
submitted to constraints coming from CMB analysis which requires
$\rho_{\tot}(\lambda)$ to be less than about $10^{-8} m_{\Pl}^4$
on the scale of the present Hubble radius \cite{Turner}.
This, in turns, converts in a upper limit on the value of $M$, $M
\lesssim 10^{-2} m_{\Pl}$. (One must impose also that $M \gtrsim
1\GeV$, so that the predictions of BBN are not spoiled
\cite{Turner}.)
Since the temperature at the end of inflation is $T_{\rm end} =
M$, we conclude that after inflation the universe is a good
conductor, $\sigma_c \gg H$, and the magnetic field evolves
adiabatically, irrespective of when it (eventually) reenters the
horizon.

\subsection{D. Form of NLE Lagrangian}

In this paragraph, we consider three models of nonlinear
electromagnetic Lagrangian. In all cases, the Lagrangian depends
on a free mass parameter, $m$, such that in the formal limit $m
\rightarrow \infty$ we recover the standard Maxwell theory. More
precisely, it results $\mathcal{L}(F) \simeq -F$ for $|F| \ll
m^4$. In the case of small fields, $|F| \lesssim m^4$,
inflation-produced fields are vanishingly small, and then cannot
explain the presently-observed fields. For this reason, we shall
restrict our analysis to the case of strong fields, $|F| \gtrsim
m$.

As a first model, we consider the family of Lagrangians
\begin{equation}
\label{toy1} \mathcal{L}(F) = -F + \sum_{i = 2}^n c_i F^i,
\end{equation}
where $i$ takes values on the integers, and the coefficients $c_j$
have dimension $[\mbox{Mass}]^{4(1-j)}$. We assume that $c_j =
m^{4(1-j)} d_j$, where $d_j$ are dimensionless constants of order
unity.
In a cosmological context, this type of Lagrangian for $n=2$ has
been widely studied in the literature (see, e.g.,
Ref.~\cite{NLE,Novello}). In Ref.~\cite{Gibbons2}, it has been
shown that the Lagrangian
\footnote{The full Lagrangian is $\mathcal{L}^{\rm KK}(F,G) = -F +
\Upsilon [(b-1) F^2 - 3G^2/2$], where $G = \frac14 F_{\mu \nu}
\widetilde{F}^{\mu \nu} = {\textbf E} \cdot {\textbf B}$, and
$\widetilde{F}^{\mu \nu} = (1/2\sqrt{-g}\,)\, \epsilon^{\mu \nu
\rho \sigma} F_{\rho \sigma}$ is the dual electromagnetic field
strength tensor. However, in this paper, we are concerned only
with nonlinear theories depending on the invariant $F$.}
\begin{equation}
\label{KK} \mathcal{L}^{\rm KK}(F) = -F + \Upsilon \, (b-1) F^2,
\end{equation}
$\Upsilon$ being a parameter with dimension $[\mbox{Mass}]^{-4}$
and $b$ a dimensionless parameter, derives from higher-curvature
gravity in Kaluza-Klein theory.
\\
In the limit of strong electromagnetic fields, we have
$\mathcal{L}(F) \simeq c_n F^n$. In this case,
Eq.~(\ref{solution}) gives
\begin{equation}
\label{Max5} {\textbf E}^2 \simeq {\textbf E}^2_1
\left(\frac{a}{a_1}\right)^{\!\!-4/(2n-1)}\!,
\end{equation}
where $a_1 = a(t_1)$, and as initial value for the electric field
we have taken that at the horizon crossing, ${\textbf E}^2_1 =
{\textbf E}^2|_{t_1}$.
From the first Bianchi identity we get $a^2 {\textbf B} \sim
(k\eta) a^2 {\textbf E} + \mbox{const}$, while from
Eq.~(\ref{Max5}) we have $a^2{\textbf E} \sim (k\eta)^{\xi}$ with
$\xi \equiv 4s(n-1)/(2n-1)$. Since we are assuming $|k\eta| \ll
1$, if $\xi<-1$ the magnetic field evolves as
\begin{equation}
\label{evolB} {\textbf B}^2 \simeq {\textbf B}^2_1
\left(\frac{a}{a_1}\right)^{\!2(2n-1-2s)/s(2n-1)}
\end{equation}
while, if $\xi \geq -1$, it scales adiabatically. For $n=1$ we
have $\xi=0$ while for $n \geq 2$ it results $\xi < -1$. Moreover,
since $2(2n-1-2s)/s(2n-1) > -4$ for $n \geq 2$, a ``superadiabatic
amplification'' ({\it i.e.} ${\textbf B}^2$ evolves less slowly
than the usual $a^{-4}$) occurs during inflation.

We now consider a ``toy model'' described by the Lagrangian
\begin{equation}
\label{toy2} \mathcal{L}(F) = -F e^{-c F},
\end{equation}
where $c = m^{-4} d$, with $d>0$ a dimensionless constant of order
unity.
The exponential self-coupling in Lagrangian (\ref{toy2}) resembles
to the exponential coupling $\propto F_{\mu \nu} F^{\mu \nu}
e^{\alpha \phi}$, $\alpha$ being a dimensional constant, between
the inflaton (dilaton) $\phi$ and the electromagnetic field,
introduced in Ref.~\cite{Ratra} (Ref.~\cite{Lemoine}). In our
case, the scalar field is replaced by the scalar $F$.
\\
In the limit of strong fields, the solution of
Eq.~(\ref{solution}) is approximatively given by
${\textbf E}^2 \simeq {\textbf E}^2_1 + m^4\ln(a_1/a)^{4/d}$.
Neglecting the logarithmic term, we have that the second model is
equivalent to the first one with $n \rightarrow \infty$.

As a third model, we consider the Born-Infeld (BI) Lagrangian
\footnote{The full Lagrangian is $\mathcal{L}^{\rm BI}(F,G) \!=\!
m^{4}[1- \sqrt{1 \!+\! 2F/m^4 \!- \!G^2/m^8}]$ (see also footnote
2).}
\begin{equation}
\label{toy3} \mathcal{L}^{\rm BI}(F) = m^{4} \! \left( 1 - \sqrt{1
+ \frac{2F}{m^4}} \, \right)\!,
\end{equation}
where, for all field configurations, the condition $2F/m^4 \geq
-1$ has to be satisfied.
Born and Infeld proposed their theory \cite{Born-Infeld} in order
to eliminate the divergence in the energy of a point-charge
particle. Indeed, in this theory the self-energy of a point-like
charge is always finite and proportional to the parameter $m$.
Born-Infeld model also appears in quantized string theory
\cite{String} (in this case, $m^2 = 2\pi\alpha'$, where $\alpha'$
is the string tension parameter). Cosmological effects of BI
electrodynamics, such as generation of an inflationary phase
\cite{NLEinfl}, have been deeply studied in recent years.
\\
Equation (\ref{solution}) gives
${\textbf E}^2 = m^4 \, [(a/a_1)^4 (m^4/{\textbf E}^2_1
-1)+1]^{-1}$.
If ${\textbf E}^2_1 < m^4$, then for $a \gg a_1$ we get the usual
behavior ${\textbf E}^2 \propto a^{-4}$. If ${\textbf E}^2_1 =
m^4$, we have ${\textbf E}^2 = m^4$ for all times. Therefore, this
model is equivalent to the first one with $n \rightarrow \infty$,
and with the condition $|F| \gtrsim m^4$, that is ${\textbf E}_1^2
\gtrsim m^4$, replaced by ${\textbf E}_1^2 = m^4$.

\subsection{E. Present Magnetic Fields}

We now derive the actual strength of magnetic fields generated
during inflation in nonlinear theories described by the above
three Lagrangians.

During de Sitter inflation any weak field, $|F| \lesssim m^4$, is
exponentially washed out. We then analyze the case in which
electric fields remain strong from the first horizon crossing to
the end of inflation. Observing that the electric field is a
non-increasing function of time, we then assume that the quantity
${\textbf E}^2(\lambda)/m^4|_{t_{\rm end}}$
is greater then 1, where $t_{\rm end}$ is the time corresponding
to the end of inflation. Taking into account Eqs.~(\ref{BE1}) and
(\ref{Max5}), we can write the above condition as
\begin{equation}
\label{chid1} m \lesssim 10^{20} \, (10^{24} \lambdakpc)^{-\mu}
\left( \frac{M}{m_{\Pl}} \right)^{\!2-\mu} \GeV,
\end{equation}
where $\mu = 1/(2n-1)$, and we used
$a_{\rm end}/a_1 = e^{N(\lambda)} \simeq 10^{24} \lambdakpc
M/m_{\Pl}$,
$N(\lambda)$ being the number of $e$-folds elapsing from the
crossing of a comoving length $\lambda$ outside the horizon to the
end of inflation \cite{Turner}.
\\
The actual value of the magnetic field follows from
Eq.~(\ref{evolB}) and is
$B_{\rm NLE} \simeq B_1 e^{-\beta N(\lambda)} (a_{\rm end}/a_0)^2$
where $B_1$, the magnetic field strength at the horizon crossing,
is given by Eq.~(\ref{BE1}), and $\beta = 1+2\mu$. The last term
in the above equation takes into account the adiabatic dilution of
the magnetic field from the end of inflation until today, $a=a_0$.
Using the relation (valid during radiation and matter dominated
eras) $a \propto g_{*S}^{-1/3} T^{-1}$, $g_{*S}(T)$ being the
number of effectively massless degrees of freedom referring to the
entropy density of the universe \cite{Kolb}, we arrive to
\begin{equation}
\label{rd1} B_{\rm NLE} \simeq 10^{-4} \, (10^{24}
\lambdakpc)^{-\beta} \left( \frac{M}{m_{\Pl}} \right)^{\!2-\beta}
\G,
\end{equation}
where we used the values $T_0 \simeq 2.35 \times 10^{-13} \GeV$,
$g_{*S}(T_0) \simeq 3.91$, and we assumed $g_{*S}(T_{\rm end})
\sim 10^2$ \cite{Kolb}.
(It is useful to know that $1\G \simeq 6.9 \times 10^{-20}
\GeV^2$.) Observe that the magnetic field during de Sitter
inflation evolves as $B \propto a^{-\beta}$. The case of standard
electromagnetic Lagrangian corresponds to $\beta = 2$. Therefore,
one finds for $\lambda = 10 \kpc$ the vanishingly small value $B
\simeq 10^{-52} \G$. In nonlinear electrodynamics, taking $M
\simeq 10^{-2} m_{\Pl}$ and $\lambda = 10 \kpc$, we find that
$B_{\rm NLE}$ is an increasing function of $n$. In particular, we
get $B_{\rm NLE} \simeq 10^{-45} \G$ for $n=2$, $B_{\rm NLE}
\simeq 10^{-33} \G$ for $n=8$, and $B_{\rm NLE} \simeq 10^{-30}
\G$ for $n \rightarrow \infty$, together with the conditions $m
\lesssim 10^8 \GeV$, $m \lesssim 10^{14} \GeV$, and $m \lesssim
10^{16} \GeV$, respectively.
\\
From the above results, we see that NLE effects are able, in
principle, to produce magnetic fields that can seed galactic
dynamo.

In the case of power-law inflation, a comoving length $\lambda$
crosses outside the horizon when \cite{Turner}
\begin{equation}
\label{outside} \rho_{\tot}(\lambda)|_{t_1} \simeq \left( 10^{24}
\, \frac{M}{m_{\Pl}} \, \lambdakpc \right)^{\!\!-2x} M^4,
\end{equation}
where $M^4 \equiv \rho_{\tot}(\lambda)|_{\rm end}$ is the total
energy density at the end of inflation, and $x \equiv
3(1+\gamma)/(1+3\gamma)$. The bound on graviton production
previously discussed translates to \cite{Turner} $x \geq x_{\m}$,
where
$x_{\m} \simeq [4 + 2 \log_{10} (M/m_{\Pl})]/[29.8 + \log_{10}
(M/m_{\Pl})]$.
In the following, we assume for simplicity that $x=x_{\m}$, and
that the electric field remains strong during inflation. This
corresponds to the ``best case scenario'', or to the minimum
dilution of the magnetic field during inflation. Taking into
account Eqs.~(\ref{BE1}) and (\ref{Max5}), and using $a_{\rm
end}/a_1 = (M^4/\rho_{\tot}(\lambda)|_{t_1})^{-3(1+\gamma)}$, the
condition
${\textbf E}^2(\lambda)/m^4|_{t_{\rm end}} \gtrsim 1$
translates into Eq.~(\ref{chid1}) with the replacement $\mu
\rightarrow \mu' = \mu + (1-\mu) x_{\m}$.
The actual strength of the inflation-produced magnetic field
follows from Eqs.~(\ref{BE1}), (\ref{evolB}), (\ref{outside}), and
is given by Eq.~(\ref{rd1}) with $\beta$ replaced by $\beta' = 1 +
2\mu'$.
The standard case of linear electrodynamics corresponds to $\beta'
= 2$, and then reduces to that studied for de Sitter inflation.
\\
In Table 1, we show the values of $B_{\rm NLE}$ (for the case of
power-law inflation) for different values of $n$ and $M$, together
with the condition on $m$ in the order that electromagnetic fields
be strong. Looking at the Table, we see that magnetic fields able
to seed galactic dynamo or directly explain galactic magnetism can
be produced.


\begin{table}
\caption{Actual strength of the magnetic field, $B_{\rm NLE}$,
produced during power-law inflation {\it via} nonlinear
electrodynamic effects at the comoving scale $\lambda = 10 \kpc$.
The last two rows refer to $\lambda = 1 \Mpc$. The index $n$ and
the mass $m$ define a particular choice of the nonlinear
Lagrangian (see text for details), while $M^4$ is the total energy
density at the end of inflation.  The magnetic field depends on
the comoving scale $\lambda$ as $B_{\rm NLE}(\lambda) \propto
\lambda^{-\beta'}$.}

\vspace{0.5cm}

\begin{tabular}{lllllll}

\hline \hline

&$n$      &$~~~M (\GeV)$   &$~~~B_{\rm NLE}(\G)$ &$~~~m (\GeV) \lesssim$ &$~~~~\beta'$ \\

\hline

&$2$      &$~~~~~~10^{16}$ &$~~~~~10^{-43}$      &$~~~~~~10^{8}$         &$~~~1.6$     \\
&$2$      &$~~~~~~10^{12}$ &$~~~~~10^{-37}$      &$~~~~~~10^{5}$         &$~~~1.1$     \\
&$2$      &$~~~~~~10^{9}$  &$~~~~~10^{-32}$      &$~~~~~~10^{3}$         &$~~~0.6$     \\
&$3$      &$~~~~~~10^{16}$ &$~~~~~10^{-37}$      &$~~~~~~10^{11}$        &$~~~1.3$     \\
&$3$      &$~~~~~~10^{12}$ &$~~~~~10^{-30}$      &$~~~~~~10^{8}$         &$~~~0.7$     \\
&$3$      &$~~~~~~10^{9} $ &$~~~~~10^{-26}$      &$~~~~~~10^{6}$         &$~~~0.1$     \\
&$3$      &$~~~~~~10^{6} $ &$~~~~~10^{-23}$      &$~~~~~~10^{3}$         &$\:-0.7$     \\
&$\infty$ &$~~~~~~10^{16}$ &$~~~~~10^{-28}$      &$~~~~~~10^{15}$        &$~~~0.9$     \\
&$\infty$ &$~~~~~~10^{12}$ &$~~~~~10^{-20}$      &$~~~~~~10^{13}$        &$~~~0.1$     \\
&$\infty$ &$~~~~~~10^{9}$  &$~~~~~10^{-14}$      &$~~~~~~10^{13}$        &$\:-0.6$     \\
&$\infty$ &$~~~~~~10^{8}$  &$~~~~~10^{-12}$      &$~~~~~~10^{12}$        &$\:-0.9$     \\

\hline \hline

\end{tabular}
\end{table}


\section{III. Conclusions}

Large-scale magnetic fields are ubiquitous in the present
universe. Astrophysical observations have proved the existence of
microgauss magnetic fields in all types of galaxies (spiral,
elliptical, barred and irregular). Remarkably, nanogauss fields
have been detected in galaxy clusters, and probably in
superclusters, with correlation lengths of $\sim 1\Mpc$.
This sort of ``cosmic magnetism'' could have been arouse out of
quantum electromagnetic fluctuations excited during an
inflationary epoch of the universe.
However, in standard, conformally-invariant theory of
electrodynamics, inflation-produced fields are vanishingly small,
and then cannot explain the presently-observed fields.
Nevertheless, a lot (sometimes exotic) mechanisms able to break
conformal invariance of Maxwell's electrodynamics, and then to
produce astrophysically interesting fields, have been proposed in
the literature.

In this paper, we have investigated the possibility to generate
magnetic fields during inflation era in nonlinear theories of
electrodynamics, in which conformal invariance is naturally
broken.
We have found that, for a wide range of parameter space of
inflationary models, magnetic fields of cosmological interest can
be created.
In particular, we have shown that magnetic fields able to seed
galactic dynamo or to explain directly the galactic magnetism
could be a natural consequence of such theories.
However, since our results strongly depend on the actual form of
the (unknown) nonlinear electromagnetic Lagrangian, they cannot
give a definitive answer to the question ``Why is our universe
magnetized?''


\vspace*{0.3cm}

\begin{acknowledgments}
L.C. would like to thank Z.~Berezhiani, D.~Comelli, M.~Gasperini,
E. Lisi, and F.~L.~Villante for useful discussions.
\end{acknowledgments}


\vspace*{0.5cm}

{\textbf{Note Added}. {\it After completion of this paper, we
noted a recent preprint by K.E. Kunze [arXiv:0710.2435], who
exploits a very similar idea, but using a different (power-law)
parametrization for the nonlinear lagrangian. We agree with
Kunze's results in the common subcase of integer-exponent
monomial. We note, however, that we can obtain cosmic magnetic
fields of the observed size via power-law inflation (see Tab. I
and related comments), without necessarily invoking a galactic
dynamo as in Kunze's paper.}


\end{document}